# Hydrogen Enhanced Intergranular Cracking – A Phenomenological Multiscale Constitutive Model

M. Amir Siddiq[a,*], Salah Rahimi[b]

[a]*School of Engineering, University of Aberdeen, AB24 3UE, United Kingdom*
[b]*Advanced Forming Research Centre, 85 Inchinnan Drive, PA4 9LJ, United Kingdom*

**Abstract**

Hydrogen enhanced cracking is one of the many failure mechanisms in metals depending on the corrosive environment. In the presented work, a multiscale constitutive model has been presented for hydrogen enhanced intergranular cracking in metals. The proposed constitutive model takes into account one of the many micromechanisms, i.e. Slip based micromechanisms active in the grain interior using crystal plasticity theory while sensitized grain boundary zone has been modelled by proposing a phenomenological constitutive model motivated from cohesive zone modelling approach. Model has been assessed through experimental data available in the literature and shows a good agreement. Model is able to capture the effects found during experiments.

*Keywords*: Hydrogen enhanced cracking; Crystal plasticity theory; Multiscale constitutive model, Intergranular failure

## 1. Introduction

Hydrogen is potentially being used as a fuel in many applications including automobiles and oil &gas industry. Whether being stored in containers, transported through pipelines or used in components such as fuel cells, it is very important to understand adverse effect of hydrogen on such components. The major adverse effect of hydrogen is found to be the reduction in the mechanical properties of metals followed by a ductile intragranular to relatively brittle intergranular cracking [1].

Experimental and computational studies have been performed in the past to understand effect and interaction of hydrogen with *dislocations* [2], [3], *precipitate-metal interface* [4], [5], *grain boundaries* [6], [7], *microvoids inside the grain* [8], [9], *other cavities and vacancies* [10], [11], *interphase boundaries* [10], [12], *deformation induced transformations, i.e. change in crystal structure or vice versa* [6], [10], *interatomic cohesive force* [13], *metal's surface energy for crack growth* [11], *brittle hydride formation at specific temperatures and ample amount of time for stable hydride formation* [14], [15]

As discussed above, there are number of experimental and modelling efforts made to understand the role of hydrogen in different metals and activation of different mechanisms have been an issue of contention for many years [3]. It has been reported that one or more of the above can play an important role in hydrogen assisted cracking. The overall aim of the current work at Multiscale mechanics of materials and manufacturing processes lab is to develop a constitutive model which takes into account all known physical mechanisms discussed above with only activating the mechanisms which have significant impact on material properties and behaviour for a specific material. The presented work is based on one of the above physical mechanism, i.e. effect of hydrogen on grain boundaries especially in the context of hydrogen assisted intergranular failure.

* Corresponding author. Tel.: +44 1224 27 2512
*E-mail address:* amir.siddiq@abdn.ac.uk

## 2. Constitutive Model

A brief outline of the constitutive model has been presented in the following while a detailed description of the model and its implementation will be reported later in a full journal article.

The grain structure is assumed to be truncated octahedron or Kelvin Cell with 14 faces comprising of 6 hexagons and 6 tetragons. Each individual grain is considered as a two phase material consisting of grain interior and the grain boundary zone as a thin shell (< 1μm) encompassing the grain interior. Grain interior is modelled using crystal plasticity theory [16]–[19] while grain boundary zone is modelled using a phenomenological constitutive model motivated from cohesive zone modelling approaches (please see references there in (Siddiq et al. 2007; Siddiq et al. 2008)). Schematic of the constitutive model is shown in Figure 1.

In the following we only present details of the calculation of overall response of the grain and a brief description of the models for both grain interior and grain boundary zone.

The stress conjugate to strain before fracture initiation is given by

$$\boldsymbol{\sigma} = \xi \boldsymbol{\sigma}_{gi} + (1 - \xi)\boldsymbol{\sigma}_{gb} \tag{1}$$

Where ξ is the volume fraction of the grain interior region and is given by

$$\xi = \frac{V_{gi}}{V_{grain}} \tag{2}$$

Where $\boldsymbol{\sigma}_{gi}$ and $\boldsymbol{\sigma}_{gb}$ are the stresses in the grain interior and grain boundary zone. $V_{gi}$ and $V_{gb}$ are volume fractions of grain interior and grain boundary zone, respectively.

Each phase, i.e. grain interior and grain boundary zone are assumed to have same deformation gradient

$$\boldsymbol{F} = \boldsymbol{F}_{gb} = \boldsymbol{F}_{gi} \tag{3}$$

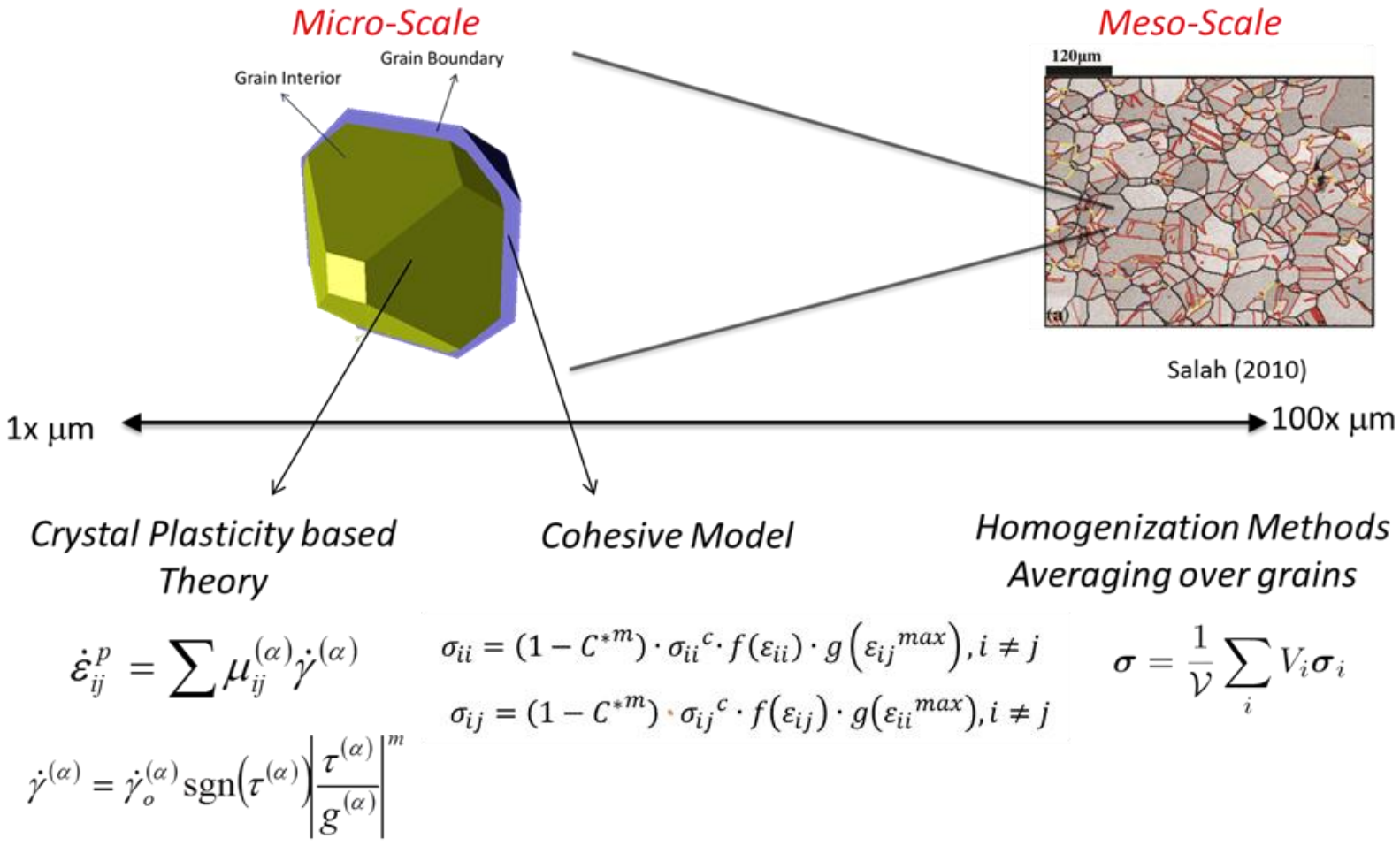

Figure 1. Schematic of Multiscale constitutive model showing how scales are bridged

Once the crack is initiated stresses are computed using the following relation

$$\sigma_{gi} = \xi \cdot e^{-\varsigma \cdot tanh\left(\frac{\varepsilon_{ij}-\left(1-C^{*m}\right)\varepsilon_{ij,1}}{\left(1-C^{*m}\right)\left(\varepsilon_{0,}-\varepsilon_{ij,1}\right)}\right)^2} \cdot \sigma_{cp} \quad (4)$$

$\sigma_{cp}$ is the stress computed through crystal plasticity theory [17], [18], [21], [22]. Where $C^*$ is a dimensionless concentration parameter and is given by

$$C_{gi}{}^{*} = \frac{(C - C_0)}{(C_E - C_0)}$$

$C_0$ -- Initial Impurity Concentration

$C_E$ -- Environmental Impurity Concentration

$\varepsilon_{ij,1gi}$ and $\varepsilon_{0,gi}$ are strains at damage initiation and complete failure respectively. $m_{gi}$ links directly the rise in normalised impurity concentration $C$ with drop in strength and strain to failure. Equation (4) is only active when intragranular fracture is found experimentally in the material, otherwise $\sigma_{gi} = \xi \cdot \sigma_{cp}$ for any stage of deformation. Where $\varepsilon_{ij,1}$, $\varepsilon_{0,}$, $m$, $C_0$, $C_E$ and $\varsigma$ can be estimated from experiments. $\varsigma$ controls the damage evolution. Polycrystalline response is computed using classical Taylor type averaging method. The homogenized stress response is computed through arithmetic averaging of overall grains.

$$\bar{\boldsymbol{F}} = \boldsymbol{F}_1 = \ldots = \boldsymbol{F}_n \qquad \bar{\boldsymbol{P}} = \frac{1}{\mathcal{V}} \sum_{i=1}^{N} V_i \boldsymbol{P}_i \quad (5)$$

Constitutive model has the capability to use microstructure character distribution data obtained by EBSD such as mean grain size and the grain orientation relationships in the form of Euler angles.
Grain interior is modelled using the following hardening model [17], [18], [21], [23]

$$h_{\alpha\alpha} = \left\{ (h_o - h_s) \sec h^2 \left[ \frac{(h_o - h_s)\gamma^{\alpha}}{\tau_s - \tau_o} \right] + h_s \right\}$$

$$h_{\alpha\beta} = q h_{\alpha\alpha} \qquad (\alpha \neq \beta) \quad (6)$$

The constitutive relation for grain boundary zone is given in Figure 1. In the following, we present the list of expressions for the constitutive model without going into the detail of derivations.
The relationships for $f(\varepsilon_{ij})$ and $g(\varepsilon_{ij})$ are given below.
Before grain boundary crack initiation the following equation is dominant

$$f(\varepsilon_{ij}) = 2\left(\frac{\varepsilon_{ij}}{\left(1 - C^{*m}\right) \cdot \varepsilon_{ij,1}}\right) - \left(\frac{\varepsilon_{ij}}{\left(1 - C^{*m}\right) \cdot \varepsilon_{ij,1}}\right)^2 \quad (7)$$

While instantly after grain boundary crack initiation the following relationships are dominant.

$$f(\varepsilon_{ij}) = e^{-\varsigma \cdot tanh\left(\frac{\varepsilon_{ij}-\left(1-C^{*m}\right)\cdot\varepsilon_{ij,1}}{\left(1-C^{*m}\right)\cdot\left(\varepsilon_0-\varepsilon_{ij,1}\right)}\right)^2} \qquad \varepsilon_{ij,1} < \varepsilon_{ij} < \varepsilon_0$$

$$g(\varepsilon_{ij}) = 2\left(\frac{\varepsilon_{ij}}{\left(1-C^{*m}\right)\cdot\varepsilon_0}\right)^3 - 3\left(\frac{\varepsilon_{ij}}{\left(1-C^{*m}\right)\cdot\varepsilon_0}\right)^2 + 1 \quad (8)$$

Where

$\varepsilon_{ij,1} = Strain\ at\ \boldsymbol{\sigma}_c$ $\qquad \varepsilon_0 = \varepsilon_{N0} = Normal\ strain\ at\ complete\ failure$

$\varepsilon_0 = \varepsilon_{T0} = Shear\ strain\ at\ complete\ failure$

And $C^*$ is dimensionless concentration in grain boundary zone while $m$ links directly the rise in normalised impurity concentration with drop in strength and strain to failure in grain boundary zone.

All simulations presented in the following are based on the code developed in the research group using FORTRAN language and all the simulations are based on local model. Total number of grains used in the model is 200. Grain misorientation was randomly distributed using experimental EBSD data.

# 3. Results - Model Assessment

Model assessment has been presented in the following by performing uniaxial tension test simulations similar to the experiments [1], [3] performed on nickel-201 with different concentrations of hydrogen charging. It must be noted that the current model only incorporates the intergranular failure mechanisms, while in the real experiments other physical mechanisms including intragranular failure due to void growth and coalescence and twinning were also present. Therefore, a qualitative comparison has been presented rather than a quantitative comparison.

Figure 2 shows the plot of stress strain response (left) and amount of intergranular fracture (right) as a function of strain and fraction of resistant grain boundary. It can be inferred from the Figure 2 that as the fraction of resistant grain boundary increases the resistance of the material also increases, i.e. stress and strain to failure initiation increases for the same amount of hydrogen charging (1012.5appm). Model is also able to capture different rates of crack propagation with changing fractions of resistant grain boundaries, as shown in Figure 2 (right). It can also be inferred from Figure 2 (right) that higher fraction of resistant grain boundaries contribute more towards crack resistance as compared to low fraction of resistant grain boundaries. The model is also able to capture the interaction of crack with resistant grain boundaries, i.e. the stress strain response increase as soon as crack reaches the resistant grain boundary, as shown in figure 2 (left).

A qualitative comparison between experiments and current model for the amount of intergranular fracture as a function of hydrogen concentration is presented in Figure 3. Experimental results are shown in Figure 3 (left). It can be inferred from Figure 3 (left) that as the hydrogen concentration is increased, the amount of intergranular fracture increases and higher fraction of resistant grain boundary leads to small amount of intergranular failure. Also, the change in the amount of intergranular fracture with respect to the hydrogen concentration increases as the fraction of resistant boundary increases. Figure 3 (right) shows the plot of the simulations performed using the presented constitutive model, it can be inferred from the figure that as the fraction of resistant grain boundary is increased the intergranular failure decreased. Also, the change in the amount of intergranular fracture with respect to the hydrogen concentration increased with the increasing fraction of resistant grain boundaries. Model is also able to capture the effect of fraction of resistant grain boundaries on amount of intergranular failure.

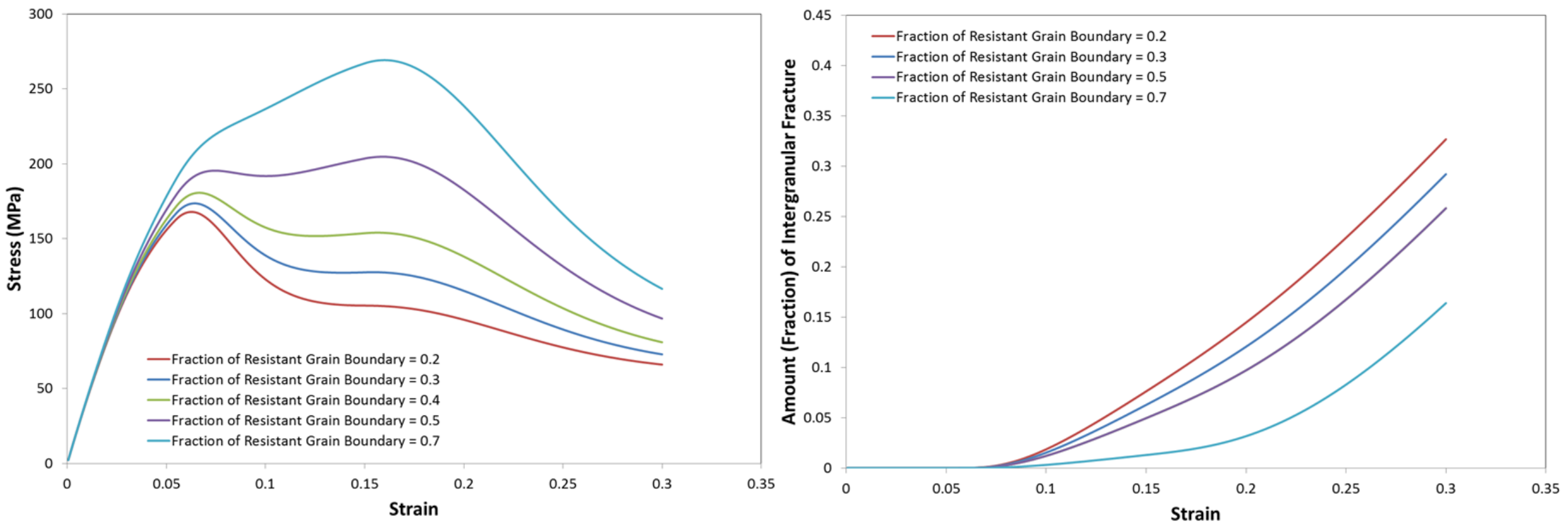

Figure 2: Stress Strain Response for different fractions of resistant (special) grain boundaries (left); Amount of intergranular fracture for different fractions of resistant boundaries (right)

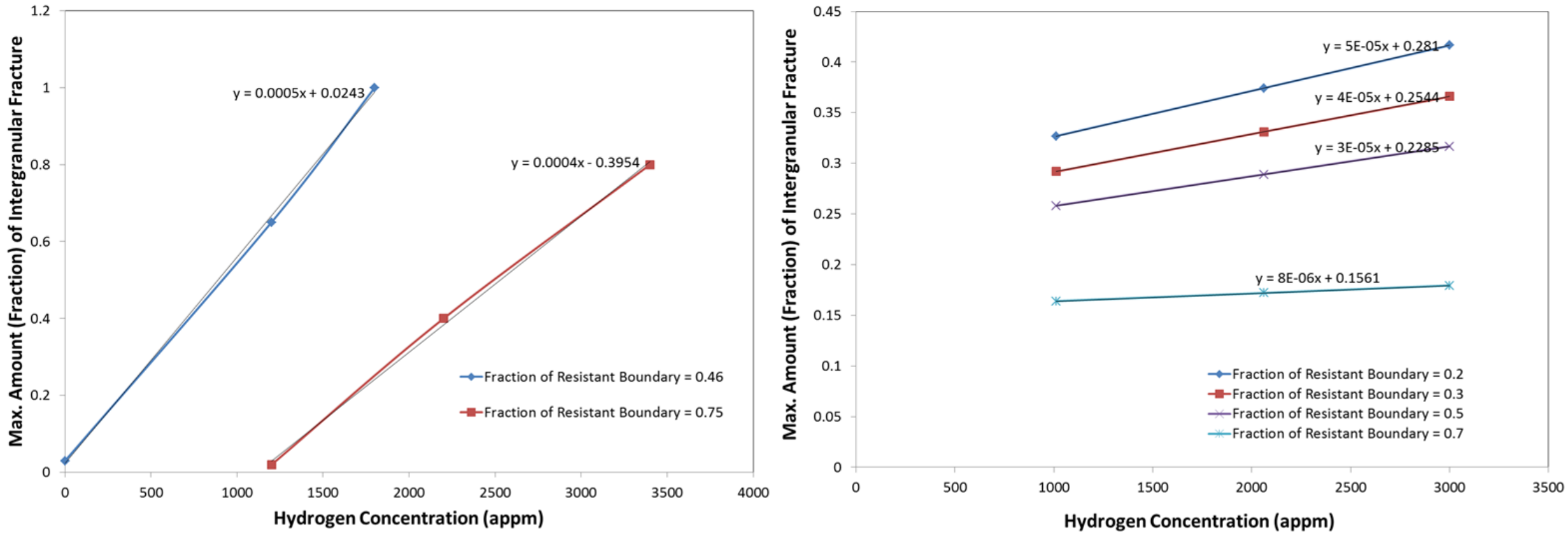

Figure 3: Amount of Intergranular fracture as a function of hydrogen concentration: Experiments (Bechtle et al 2009)[left], Simulation [right]

## 4. Conclusion

A Multiscale constitutive model for hydrogen enhanced intergranular failure in metals was presented in this work. Model takes into account the grain size, grain orientation distribution using crystal plasticity theory. Grain boundary zone is modelled using a constitutive model motivated from cohesive model. Constitutive model has the capability of using EBSD data to define grain size, orientation and distribution of grain boundary types, i.e. resistant and susceptible grain boundaries. Constitutive model shows a good qualitative agreement with experimental data.

As a future outlook, the model is currently being extended to account for trans-granular failure due to void growth and coalescence along with twinning as another plasticity mechanism.

## Acknowledgements

Authors would like to acknowledge Dr. Rahul Bhattacharya (Advanced Forming Research Centre) for useful discussions and motivating remarks throughout this work.